\documentclass[prb,aps]{revtex4}
\usepackage{graphicx}
\begin{document}
\title{Nonlinear Effect of Transport Current on Response
of Metals to Electromagnetic Radiation}
\author{S.A. Derev'anko, G.B. Tkachev, V.A. Yampol'skii}

\affiliation{Institute for Radiophysics and Electronics, NASU \\
12 Acad. Proscura St., 310085 Kharkov, Ukraine\\ E-mail:
stanislv@ire.kharkov.ua}
\begin{abstract}

The nonlinear interaction of DC current flowing in a thin metal
film with an external low-frequency AC electromagnetic field is
studied theoretically. The nonlinearity is related to the
influence of the magnetic field of the DC current and the magnetic
field of the wave on the form of electron trajectories. This
magnetodynamic mechanism of nonlinearity is the most typical for
pure metals at low temperatures. We find that such interaction
causes sharp kinks in the temporal dependence of the AC electric
field of the wave on surface of the sample. The phenomenon of
amplification of the electromagnetic signal on the metal surface
is predicted. We also calculate the nonlinear surface impedance
and show that it turns out to be imaginary-valued and its modulus
decreases drastically with the increase of the wave amplitude.

\end{abstract}
\maketitle

\section{Introduction}
As well-known metals possess quite peculiar nonlinear
electrodynamic properties (see., for example, Refs.~\cite{rev1,rev2}).
Indeed, nonlinearity in a response of plasma or semiconductors to
an electromagnetic excitation is usually associated to significant
deviation of the electron subsystem from equilibrium. To the contrary,
nonequilibrium in metals as a rule is weak due to large
concentration of charge carriers. Nevertheless, a nonlinear regime in
these media is rather easy to observe. Such a situation is possible
because of the fact that nonlinearity in metals does not necessarily related
to the nonequilibrium of the electron subsystem. Nonequilibrium is
caused by the existence of a weak electric field,
while nonlinearity is originates from strong magnetic fields.
The Lorentz force, determined by the magnetic
field of an electromagnetic wave or magnetic field of the transport current,
affects the dynamics of charge carriers. As a result, the conductivity of
a sample placed in an AC electromagnetic field
depends on the spatial distribution of magnetic field of the wave.
Such a {\it magnetodynamic} mechanism of nonlinearity is
inherent to pure metals at low temperatures, if the mean free path of
conducting electrons is rather large.

Magnetodynamic nonlinearity causes a number of nontrivial phenomena in the
electrodynamics of metals. As an example, one can mention generation of
the {\it current states}~\cite{Dolg,CS} in a sample placed in a DC external
magnetic field. The sample acquires a DC magnetic moment if irradiated by
an additional strong AC electromagnetic field.
 The magnitude of the magnetic moment depends in a hysteretic manner
on an external DC magnetic field.  Under current states conditions, a
hysteresis-like interaction of radiowaves is observed~\cite{int} as well
as the appearance of electromagnetic dissipative structures~\cite{diss}.
This specific mechanism of nonlinearity in metals causes a decrease of
collisionless damping of helicons~\cite{helic1}.  Therefore, the
spiral waves with large amplitudes can propagate even in conditions
when there is no their linear electromagnetic excitations~\cite{helic2}.
Magnetoplasmic shock waves~\cite{shock} and soliton-like
excitations~\cite{waves} are also predicted for the regime of strong
magnetodynamic nonlinearity.

In the present paper, we study a novel manifestation of
magnetodynamic nonlinearity, namely, interaction of an external
electromagnetic wave and a strong DC transport current in a thin metal
film, which is also displayed in a quite unusual way.
The sample of thickness $d$ is assumed to be much smaller than electron
mean free path $l$,
\begin{equation}
d\ll l,
\label{d}
\end{equation}
and electron scattering on a surface of the film is supposed to be
diffuse. It is known~\cite{Snapiro} that in the static case
(when external AC signal is absent), the magnetic field of a current
can essentially affect the conductivity of a thin metal specimen and,
thus, its current-voltage characteristics (CVC).  In this situation
the value $I$ of the current is rather small so that the typical radius
of curvature $R(I)$ of electron trajectories in magnetic field is much
greater than the film thickness,
\begin{equation}
d\ll R(I),\qquad\qquad R(I)=cp_F/eH(I)\propto I^{-1}.
\label{weak-I}
\end{equation}
Here $-e$ and $p_F$ represent electron charge and Fermi momentum,
respectively.  In Ref.~\cite{Snapiro}, it was shown that nonlinear
peculiarities of CVC are connected to the antisymmetrical spatial
distribution of the magnetic field of the DC current inside the sample.
The magnetic field equals to zero at the middle of the film and
takes the values $H$ to $-H$ at the opposite boundaries
of the sample, where
\begin{equation}
H=2\pi I/cD.
\label{H}
\end{equation}
In this formula, $c$ denotes the speed of light in vacuum and $D$ is the
sample width. Spatially alternating field of the DC current entraps
a part of electrons in a potential well. Trajectories of such particles
are flat curves winding around the plane of alternation of the
magnetic field. The relative part of the {\it trapped} electrons in
the order of magnitude is equal to the typical angle $(d/R)^{1/2}\ll 1$
of their crossing of that plane. Taking into account that trapped carriers
do not collide with the film boundaries and interact with the electric field
along their whole free path $l$, one can write the following estimating
formula for their conductivity $\sigma_{tr}$:
\begin{equation}
\sigma_{tr}\sim\sigma_0 (d/R(I))^{1/2}\propto I^{1/2}.
\label{trap-est}
\end{equation}
Here $\sigma_0$ represents the conductivity of the bulk sample.
At the same time, there exist {\it flying} electrons which do collide
with the boundaries of the specimen and, according to Ref.~\cite{Fuks}, have
the conductivity of the order of $\sigma_0 (d/l)$. Apparently in the
range of rather strong currents, when the inequality
\begin{equation}
(dR(I))^{1/2}\ll l,
\label{trap-cond}
\end{equation}
holds, the conductivity of the film is determined by the group of
the trapped carriers. As a result, we observe the deviation from
the Ohm's law: the voltage $U$ is proportional to the square root
of current,
\begin{equation}
U\propto I^{1/2}.
\label{root}
\end{equation}
For the film with thickness $d=10^{-3}$ cm, the electron mean free path
$l=10^{-1}$ cm, and Fermi momentum $p_F=10^{-19}$ g$\cdot$ cm/s
the nonlinearity becomes noticeable ($(dR)^{1/2}\sim ~l$) at values of
the magnetic field $H(I)$ about $1$ Oe. The theory developed in
Ref.~\cite{Snapiro} is in a good qualitative agreement with
experimental data (see, for example, Ref.~\cite{Fish1}).

In an external magnetic field $\bf h$, collinear to the magnetic field
of the current, the plane of the sign alternation for the magnetic field
shifts to one of the two boundaries of the film (see Fig.~1). That in turn
leads to appreciable diminution of the number of the trapped
particles and, therefore, their conductivity. In particular, such a
situation would take place under symmetrical irradiation of the film by
the low-frequency electromagnetic wave of large amplitude. The frequency
is supposed to be so small that the AC magnetic field ${\bf h}(t)$ of the
wave is virtually uniform across the metal (i.e. the wave penetration
depth $\delta$ is much greater than the sample thickness $d$).  Then
the conductivity of metal essentially depends on time and, therefore,
strong nonlinear effects in the sample response to the AC electromagnetic
excitation should appear. Being of interest from the
both theoretical and experimental points of view this problem has not
been investigated yet.

In the present paper we study theoretically the temporal
dependence of electric field at the surface of the film, which
carries a strong DC current of the fixed value $I$ satisfying
inequalities (\ref{weak-I}) and (\ref{trap-cond}). It is shown
that with an increase of the amplitude $h_m$ of the AC magnetic field this
dependence becomes anharmonic, turning into a series of sharp nonanalytic
peaks. The case of sufficiently high amplitudes $h_m>H$,
when the total magnetic field in the sample is spatially alternating
during some part of the wave period and has constant sign during the
other part, is of particular interest. In such a situation, the
electric field is also characterized by kinks in its temporal dependence
due to the periodical appearance and disappearance of the group of trapped
carriers. The effect of amplification of the electric signal on the film
surface is predicted as well. It turns out that, because of the presence of
the strong DC transport current in the sample, the absolute value of the
AC electric field of the wave is $l/(dR)^{1/2}\gg 1$ times as many as the
corresponding magnitude in the absence of the DC current.

We also calculate the nonlinear surface impedance of the film,
which turns out to be pure imaginary value in the main
approximation in the parameter $d/\delta\ll 1$, and show that its
modulus monotonically diminishes with the growth of the AC
amplitude decreasing in $l/(dR)^{1/2}\gg 1$ times. Simultaneously
the conductivity of the trapped particles falls down and,
consequently, $\delta$ increases.

\section{Problem statement and geometry}
Consider a metal film of the thickness $d$ with a DC current $I$
flowing along. The sample is irradiated from both sides by
a monochromatic electromagnetic wave with a magnetic
component collinear to the magnetic field of the current.
The $x$-axis is oriented perpendicularly to the film boundaries.
The plane $x=0$ corresponds to the middle of the sample (see Fig.~1).
\begin{figure}[btpf]
\centering \scalebox{0.5}[0.5]{\includegraphics[bb=65 29 778
562]{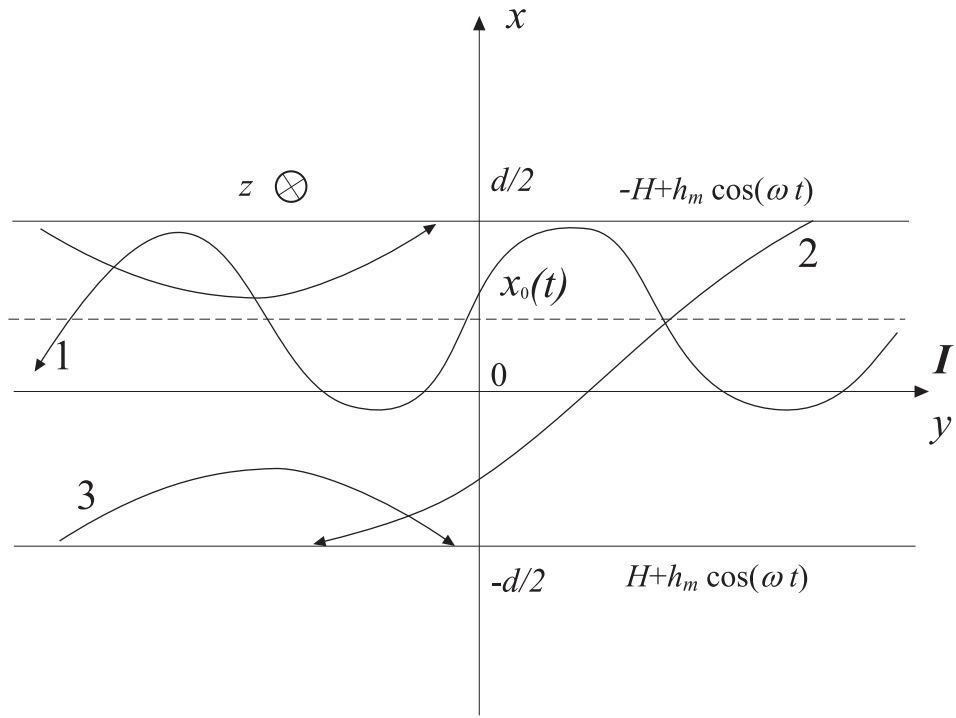}} \caption{Geometry of problem. Schematic
representation of the trajectories of trapped (1), flying (2), and
surface (3) electrons.}
\end{figure}
The $y$-axis is directed along the current,
and the $z$-axis is parallel to the vector ${\bf {\cal H}}(x,t)$ of the
total magnetic field which is a sum of the DC magnetic field of the
current ${\bf H}(x,t)$ and the AC magnetic field of the wave ${\bf
h}(x,t)$,
\begin{equation}
{\bf {\cal H}}(x,t)= \{0, 0, H(x,t)+h(x,t)\}.
\label{vec-H}
\end{equation}
The film length $L$ (the dimension along the $y$-axis) and its width $D$
($z$-axis dimension) are much greater than the sample thickness $d$. We
assume diffuse scattering of the electrons on the film boundaries.
Maxwell's equations in the assumed geometry can be written as
\begin{equation}
-\frac{\partial {\cal H}(x,t)}{\partial x}= \frac{4\pi}{c}j(x,t),\qquad
\frac{\partial E(x,t)}{\partial x}= -\frac{1}{c}\frac{\partial {\cal
H}(x,t)}{\partial t},
\label{Max}
\end{equation}
where $j(x,t)$ and $E(x,t)$ represent the $y$ components of the current
density and the electric field. Boundary conditions for Eqs.~(\ref{Max})
are
\begin{equation}
{\cal H}(\pm d/2,t)=h_m\cos\omega t\mp H.
\label{bound}
\end{equation}
Let $H$ be the absolute value of the magnetic field on the surface
of the metal film and $h_m$ denote a wave amplitude. According to
Eq.~(\ref{H}), the field $H$ is determined by the DC current $I$.
No special relation between magnitudes $H$ and $h_m$ is assumed.

We consider a quasistatic situation when the wave frequency $\omega$
is much less than the relaxation frequency $\nu$ of the charge carriers,
\begin{equation}
\omega\ll\nu.
\label{low}
\end{equation}
Here we suppose that the AC magnetic field inside the sample
is quasiuniform and virtually does not differ from its value on the sample
surface, $h(x,t)\simeq h_m\cos\omega t$. In other words, the typical
spatial scale $\delta(\omega)$ of variation of the AC magnetic field
is much greater than the film thickness $d$. Furthermore we assume that
the curvature radius $R(x,t)$ of electron trajectories in the total
magnetic field ${\cal H}(x,t)$ is also much greater than $d$,
\begin{equation}
d \ll \delta(\omega), \qquad d\ll R(x,t), \qquad
R(x,t)=cp_F/e|{\cal H}(x,t)|.
\label{quasi}
\end{equation}
%

\section{Electron dynamics, current density, and CVC of film}
Let us consider electron dynamics in the nonuniform AC magnetic field
${\cal H}(x,t)$. We shall assume the following gauge of the vector potential:
\begin{equation}
{\bf A}(x,t)=\{0,A(x,t),0\}, \qquad A(x,t)=\int^{x} dx^\prime
{\cal H}(x^\prime,t).
\label{gauge}
\end{equation}
It is suitable to choose the lower limit of integration in
Eq.~(\ref{gauge}) depending on whether or not there exists the
plane $x=x_0(t)$ of the sign alternation of the magnetic field
${\cal H}(x,t)$ at the present moment. This plane exists during
the time intervals when $h_m|\cos\omega t|<H$ because the values
$h_m\cos\omega t-H$ and $h_m\cos\omega t+H$ of the total magnetic
field at the film boundaries have opposite signs (see
Eq.~(\ref{bound})). In this case, one should take $x_0(t)$ as the
lower limit in integral (\ref{gauge}). Then the vector potential
$A(x,t)$ is negative. It reaches its maximum value (which equals
to zero) at the point $x=x_0(t)$. Within other time intervals,
when the inequality $h_m|\cos\omega t|>H$ holds, the magnetic
field ${\cal H}(x,t)$ inside the sample is of a constant sign.  In
such a situation, one should choose ${\rm sign}(\cos\omega t)d/2$
(${\rm sign}(x)$ is the sign function) as a lower limit of the
integration.  In this case, vector potential, also being negative,
vanishes at one of the boundaries of the film.

The integrals of motion of electron in the field ${\cal H}(x,t)$
are the total energy (it equals to the Fermi energy) and the
canonical momenta $p_z=m v_z$ and $p_y=m v_y-e A(x,t)/c$ (m is the
electron mass). The electron trajectory in a perpendicular to the
direction of the magnetic field plane is determined by the
velocities $v_x(x,t)$ and $v_y(x,t)$. In the case of Fermi-sphere
with radius $p_F=m v$, we obtain
\begin{equation}
 |v_x(x,t)|=(v_{\bot}^2-v_y^2)^{1/2}, \ \
v_{\bot}=(v^2-v_z^2)^{1/2}, \ \ v_y(x,t)=(p_y+e A(x,t)/c)/m.
\label{velocity}
\end{equation}

Classically allowable regions of the electron motion along the $x$ axis
are determined by the inequalities,
\begin{equation}
-p_y-m v_{\bot}\leq e A(x,t)/c\leq -p_y+m v_{\bot}.
\label{limits}
\end{equation}
These inequalities provide the positivity of the radicand in
Eq.~(\ref{velocity}) for $|v_x(x,t)|$.

The regions of the electron motion in the phase plane $(x,p_y)$
are described schematically in Fig.~2 for two cases: when there exists the
plane $x=x_0(t)$ of the sign alternation of the magnetic field ${\cal
H}(x,t)$ (Fig.~2, a) and when such plane is absent (Fig.~2, b).
\begin{figure}[btpf]
\centering \scalebox{0.7}[0.7]{\includegraphics[bb=30 147 709
469]{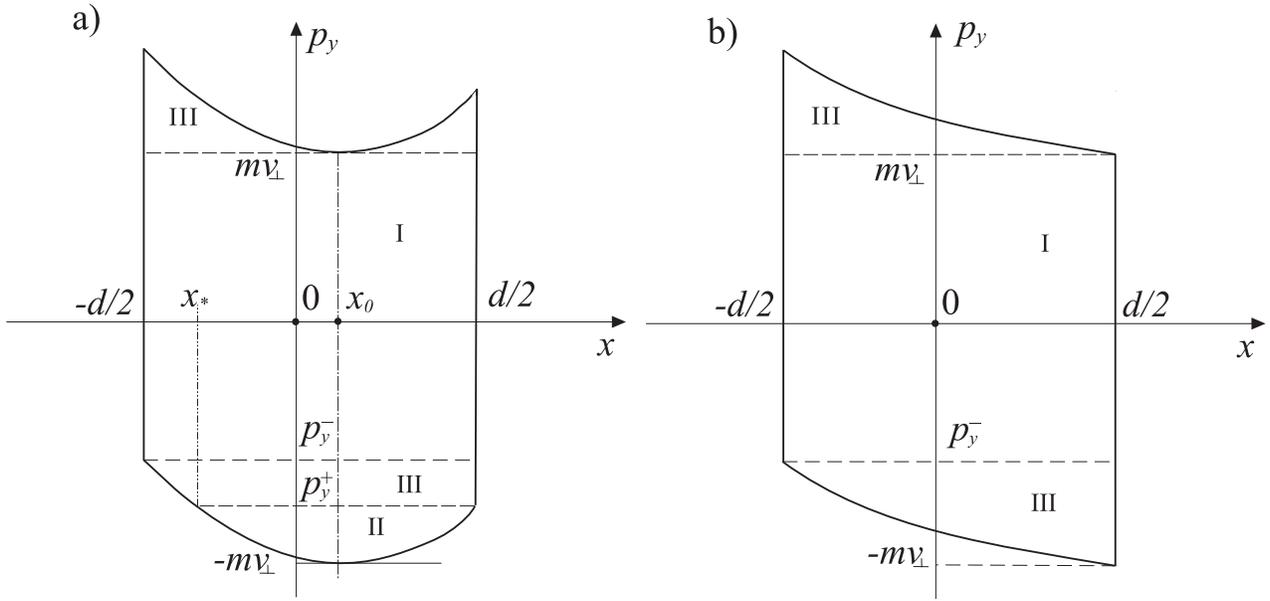}} \caption{Phase space $(p_y,x)$. Regions of
existence of flying (I), trapped (II), and surface (III) particles
in spatially sign alternating
 (a) and of constant sign (b) total magnetic field.}
\end{figure}
For definiteness we have chosen the moment of time when the
magnetic field of the wave is positive ($\cos\omega t>~0$). The
upper border on the phase plane is described by the curve
$p_y=mv_{\bot}-eA(x,t)/c$ and the lower one is given by
$p_y=-mv_{\bot}-eA(x,t)/c$. The electrons are naturally divided in
three groups depending on the sign and value of the integral of
motion $p_y$.  Below, we give inequalities determining the regions
of their existence at an arbitrary moment of time.

\begin{enumerate}

\item Flying electrons:

\begin{equation}
 p_y^-\equiv -
mv_{\bot}-eA[-{\rm sign}(\cos\omega t)d/2,t]/c\leq p_y \leq mv_{\bot},
\qquad |x|\leq d/2.
\label{flying}
\end{equation}
These particles collide with the both boundaries of the film.
Their trajectories do not twist significantly because of $d\ll R(x,t)$.
Flying electrons exist at every moment of time irrespective of
the presence of the plane $x=x_0(t)$ (i.e irrespective of the relation between
$h_m\cos\omega t$ and $H$).

\item Trapped electrons:
They appear during the periods of time when $h_m|\cos\omega t|<H$ and
the total magnetic field ${\cal H}(x,t)$ within the sample passes
trough zero.
Their states are bounded by the region (see. Fig.~2,a),
\begin{eqnarray}
&-mv_{\bot}\leq p_y\leq p_y^{+}\equiv
-mv_{\bot}-eA[{\rm sign}(\cos\omega t)d/2,t]/c,& \nonumber\\
&x_*(t){\rm sign}(\cos\omega t)<x{\rm sign}(\cos\omega t)<d/2.&
\label{trapped}
\end{eqnarray}
Here $x_*(t)$ represents the breakpoint of the trapped electron
most distant from the film boundary. One can find it
from the equation,
\begin{equation}
A(x_*,t)=A[{\rm sign}(\cos\omega t)d/2,t].
\label{x*}
\end{equation}
According to Eq.~(\ref{trapped}), this electron group occupies
the region $x_*(t)<x<~d/2$ when $\cos\omega t>0$ and the region
$-d/2<x<x_*(t)$ if $\cos\omega t<0$. The trajectories of trapped particles
are almost flat oscillating curves due to periodical motion of the particles
along $x$-direction and the uniform motion along the $y$-and $z$-axes.
The temporal period of oscillations with respect to the plane $x=x_0$
equals to $2T$, where
\begin{equation}
T=\int^{x_2(t)}_{x_1(t)}\frac{dx}{|v_x(x,t)|}.
\label{Period}
\end{equation}
The breakpoints $x_1(t)$ and $x_2(t)$ ($x_1(t)<x_0(t)<x_2(t)$)
are the roots of the equation,
\begin{equation}
eA(x_{1,2},t)/c=-mv_{\bot}-p_y.
\label{Breakpoints}
\end{equation}

\item Surface electrons:

These particles collide only with one of the boundaries of the film.
In our case of diffuse scattering of the electrons on the surface,
their influence on the nonlinear conductivity of metal is
negligible~\cite{Snapiro}. Thus, we do not take them into account thereafter.
\end{enumerate}

The current density of the flying and trapped particles can be deduced by
means of solving the Boltzmann kinetic equation. One
should linearize the kinetic equation with respect to the electric field
$E(x,t)$, which can be represented as a sum,
\begin{eqnarray}
\label{E-tot}
E(x,t)&=&E_0+{\cal E}(x,t),
\nonumber \\
{\cal E}(x,t)&=&-\frac{1}{c}\left(\frac{\partial A(x,t)}{\partial
t}- \frac{\partial \bar A(t)}{\partial t}\right).
\end{eqnarray}
Here the first term, $E_0$, is a potential (uniform) component
and ${\cal E}(x,t)$ is a rotational (nonuniform) field of the wave.
Spatial averaging of the latter over the $x$-axis direction gives zero.
The value $\bar A(t)$ represents a spatially averaged magnitude of the vector
potential,
\begin{equation}
\bar A(t)=\frac{1}{d}\int^{d/2}_{-d/2}A(x',t)dx'.
\label{mean}
\end{equation}

The magnetodynamic nonlinearity is accounted for in the kinetic equation
by means of terms which contain the total magnetic field ${\cal H}(x,t)=H(x,t)
+h(x,t)$ entering the Lorentz force. We calculate the current density
in the main approximation with respect to the small parameter
$d/\delta(\omega)$ (see Eq.~(\ref{quasi})). In this approximation, as it was
mentioned above, the AC magnetic field $h(x,t)$ becomes spatially uniform
and is equal to its boundary value, $h(x,t)=h_m\cos\omega t$. The electric
field is also independent of the $x$-coordinate and
coincides with the value $E_0(t)$. For the case of uniform electric
and external magnetic fields, the current density was obtained in
Ref.~\cite{Snapiro}. If the conditions (\ref{weak-I}) and
(\ref{trap-cond}) hold the following asymptotics for the current
density of the flying and trapped electrons are valid:
$$
j_{fl}(t)=\sigma_{fl}(t)E_0(t),
$$
\begin{equation}
\sigma_{fl}(t)=\frac{3}{8}\sigma_0\frac{d}{l}\ln\frac{R_+(t)}{d}, \quad
R_\pm(t)=cp_F/e|h_m|\cos\omega t|\pm H|,
\label{j-flying}
\end{equation}
$$
j_{tr}(x,t)=\sigma_{tr}(x,t)E_0(t),
$$
\begin{equation}
\sigma_{tr}(x,t)=\frac{36\pi^{1/2}}{5\Gamma^2(1/4)}\sigma_0
\Big\{\frac{e}{cp_F}
\left[A(x,t)-A({\rm sign}(\cos\omega
t)d/2,t)\right]\Big\}^{1/2},
\label{j-trapped}
\end{equation}
$$
x_*(t){\rm sign}(\cos\omega t)<x{\rm sign}(\cos\omega t)<d/2.
$$
In the limit $\omega\to 0$, Eqs.~(\ref{j-flying}) and
(\ref{j-trapped}) transform into the corresponding formulae of
Ref.~\cite{Snapiro}.

Let us substitute the current density in the first of Maxwell's
equations (\ref{Max}) for its asymptotic expressions (\ref{j-flying}) and
(\ref{j-trapped}) and introduce a dimensionless coordinate and vector
potential,
\begin{equation}
\xi=2x{\rm sign}(\cos\omega t)/d, \qquad a(\xi,t)=A(x,t)/A({\rm
sign}(\cos\omega t)d/2,t).
\label{xi,a}
\end{equation}
The equation for the quantity $a(\xi,t)$ has the form,
\begin{equation}
\frac{\partial^2a(\xi,t)}{\partial \xi^2}=u\cases{
r[1-a(\xi,t)]^{1/2} + 1, &$\xi_*(t)\leq\xi\leq 1$, \cr
 & \cr
1, &$-1\leq\xi\leq\xi_*(t),$
\cr}
\label{eq-a}
\end{equation}
\begin{equation}
\xi_*(t)=2x_*(t){\rm sign}(\cos\omega t)/d.
\label{xi*}
\end{equation}
The dimensionless
coordinate $\xi_*(t)$ confines the region of existence of the
trapped particles and, according to Eqs.~(\ref{x*}) and (\ref{xi,a}),
satisfies
the equation, $a(\xi_*,t)=1$. The parameter $r$ represents the ratio of
the maximum magnitude of the conductivity of the trapped electrons to
the conductivity of the flying particles,
\begin{equation}
 r=\frac{\sigma_{tr}(x_0)}{\sigma_{fl}}=\frac{96\pi^{1/2}}{5\Gamma^2(1/4)}
\frac{l}{d}\left[\frac{e}{cp_F}|A({\rm sign}(\cos\omega
t)d/2,t)|\right]^{1/2}\ln^{-1}(R_+/d).
\label{r}
\end{equation}
The dimensionless quantity, $u$, is related to the voltage $U=E_0L$
on the sample,
\begin{equation}
u=\frac{U}{cL|A({\rm sign}(\cos\omega
t)d/2,t)|/\pi\sigma_{fl}d^2}.
\label{eps}
\end{equation}
Equation (\ref{eq-a}) should be solved together with the boundary conditions,
$$
\frac{\partial a(1,t)}{\partial \xi}=
\frac{d}{2}\frac{h_m|\cos\omega t|-H}{A({\rm sign}(\cos\omega
t)d/2,t)},
$$
\begin{equation}
\frac{\partial a(-1,t)}{\partial \xi}=
\frac{d}{2}\frac{h_m|\cos\omega t|+H}{A({\rm sign}(\cos\omega
t)d/2,t)},\quad
a(1,t)=1.
\label{cond-a}
\end{equation}
The first two of these expressions are dimensionless boundary
conditions (\ref{bound}), and the third one is a consequence of
normalization (\ref{xi,a}) of the vector potential.

Within the interval $\xi_*(t)\leq\xi\leq 1$, the solution of
Eq.~(\ref{eq-a}) is symmetrical with respect to the point
$\xi_0(t)=(1+\xi_*(t))/2$, where the dimensionless vector potential reaches
its minimum value (which equals to zero, $a(\xi_0,t)=\partial
a(\xi_0,t)/\partial \xi=0$). This solution is described by the formula,
\begin{equation}
|\xi-\xi_0(t)|=(3/4ru)^{1/2}\int_{0}^{a(\xi,t)}d\zeta
[1-(1-\zeta)^{3/2}+3\zeta/2r]^{-1/2}.
\label{sol-1}
\end{equation}
One can not obtain the field distribution and the current
density within the region of existence of the trapped electrons
in an explicit form. However, by means of Eq.~(\ref{sol-1}), it is possible
to calculate the average magnitude of the conductivity of the trapped
carriers (\ref{j-trapped}) within the interval (\ref{trapped}),
\begin{eqnarray}
\frac{{\bar\sigma_{tr}}}{\sigma_{fl}}&=&
r\int_{0}^{1}d\zeta (1-\zeta)^{1/2}
[1-(1-\zeta)^{3/2}+3\zeta/2r]^{-1/2}\nonumber\\
&\times &\left(\int_{0}^{1}d\zeta [1-(1-\zeta)^{3/2}+3\zeta/2r]^{-1/2}\right)^{-1}.
\label{sr-trap}
\end{eqnarray}
The bar above $\sigma_{tr}$ denotes spatial averaging. In the remaining
region of the sample ($-1\leq\xi\leq\xi_*(t)$), there exist only
flying electrons, and the solution of Eq.~(\ref{eq-a})
is given by the following formula:
\begin{equation}
a(\xi,t)=1-(2u)^{1/2}(1+2r/3)^{1/2}(\xi-\xi_*(t))+u
(\xi-\xi_*(t))^2/2.
\label{sol-2}
\end{equation}
Expressions (\ref{sol-1}) and (\ref{sol-2}) and their derivatives
are sewn together at the point $\xi=\xi_*(t)$. The solution given by
Eqs.~(\ref{sol-1}) and (\ref{sol-2}) contains three parameters, $\xi_0$, $u$
and $r$, which should be found from boundary conditions (\ref{cond-a}).
It is essential that the value $A({\rm sign}(\cos\omega t)d/2,t)$ of
the vector potential appearing in Eq.~(\ref{cond-a}) is not an
independent parameter due to its relation to $r$ via formula (\ref{r}).

Adding term by term the first two boundary conditions in Eq.~(\ref{cond-a})
and using Eqs.~(\ref{sol-1}) and (\ref{sol-2}), and (\ref{eps}), we find the
following expression for the drift of the plane $x=x_0$:
\begin{equation}
\xi_0=2x_0{\rm sign}(\cos\omega t)/d=\frac{áLh_m|\cos\omega t|}{2\pi
U\sigma_{fl}d},\qquad h_m|\cos\omega t|\leq H.
\label{xi-0}
\end{equation}
In order to determine the value of $u$ (i.e. the voltage $U$), let us
integrate the left and right-hand sides of Eq.~(\ref{eq-a})
from -1 to 1 taking into account the boundary conditions for the
derivative $\partial a(\xi,t)/\partial\xi$ in (\ref{cond-a}). The integral
of the function $[1-a(\xi,t)]^{1/2}$ appearing in the right-hand side
can be reduced to the product $2(1-\xi_0){\bar\sigma_{tr}}/r\sigma_{fl}$
with the use of the condition $a(1,t)=1$. Taking this into consideration
as well as formulae (\ref{eps}) and (\ref{xi-0}) for the quantities
$u$ and $\xi_0$, we have after some simple transformations,
\begin{equation}
U=\frac{cL}{2\pi d\sigma_{fl}(t)}
\frac{H(I)+({\bar\sigma_{tr}}/\sigma_{fl})h_m|\cos\omega t|}
{1+{\bar\sigma_{tr}}/\sigma_{fl}},\qquad h_m|\cos\omega t|\leq H.
\label{VAX}
\end{equation}
According to Eq.~(\ref{sr-trap}), the ratio of conductivities,
${\bar\sigma_{tr}}/\sigma_{fl}$, depends on the parameter $r$.
Using expression (\ref{eps}) for $u$,
relation (\ref{r}) between $A({\rm sign}(\cos\omega t)d/2,t)$ and $r$, and
solution (\ref{sol-1}), we obtain from the first boundary condition in
Eq.~(\ref{cond-a}) the algebraic equation for $r$,
\begin{equation}
 r^2(1+2r/3)=
\left(\frac{H-h_m|\cos\omega t|}{{\tilde H}}\right)^2
\frac{{\tilde U}}{U\ln^3(R_+/d)},\qquad h_m|\cos\omega t|\leq H.
\label{eq-r}
\end{equation}
Here we have introduced the following notations:
\begin{equation}
{\tilde H}=\frac{25\Gamma^4(5/4)}{9\pi}\frac{cp_Fd}{el^2},\qquad
{\tilde U}=\frac{4clL{\tilde H}}{3\pi\sigma_0 d^2}.
\label{obozn}
\end{equation}
The parameters ${\tilde H}$ and ${\tilde U}$ represent those magnitudes of
the magnetic field and voltage for which the characteristic length
$(Rd)^{1/2}$ of the arch of electron's trajectory is of the order of
the mean free path $l$.

Expressions (\ref{sr-trap}), (\ref{VAX}), and (\ref{eq-r}) define, in an
implicit form, dependence of the voltage $U$ on the current $I$ for the
case $h_m|\cos\omega t|\leq H$. At these conditions there exists the plane
of the alternation of sign of the total magnetic field within the sample.
If the opposite
inequality, $h_m|\cos\omega t|\geq H$, is valid, the trapped electrons are
absent ($r=0$, $\xi_* =1$, $\sigma_{tr}=0$) and CVC is described by the
formula,
\begin{equation}
U=\frac{cLH(I)}{2\pi d\sigma_{fl}(t)},\qquad h_m|\cos\omega t|\geq H.
\label{VAX1}
\end{equation}

As seeing from formula (\ref{VAX}), the voltage on the
sample displays nonanalytical behavior vs. time: the dependence
$U(t)$ has kinks at the moments when the AC magnetic field $h_m\cos\omega
t$ vanishes. This is an essentially nonlinear effect caused by
the contribution of a large group of trapped electrons into the electric
current.
\begin{figure}[bt]
\centering \scalebox{0.8}[0.8]{\includegraphics[bb=156 82 506
547]{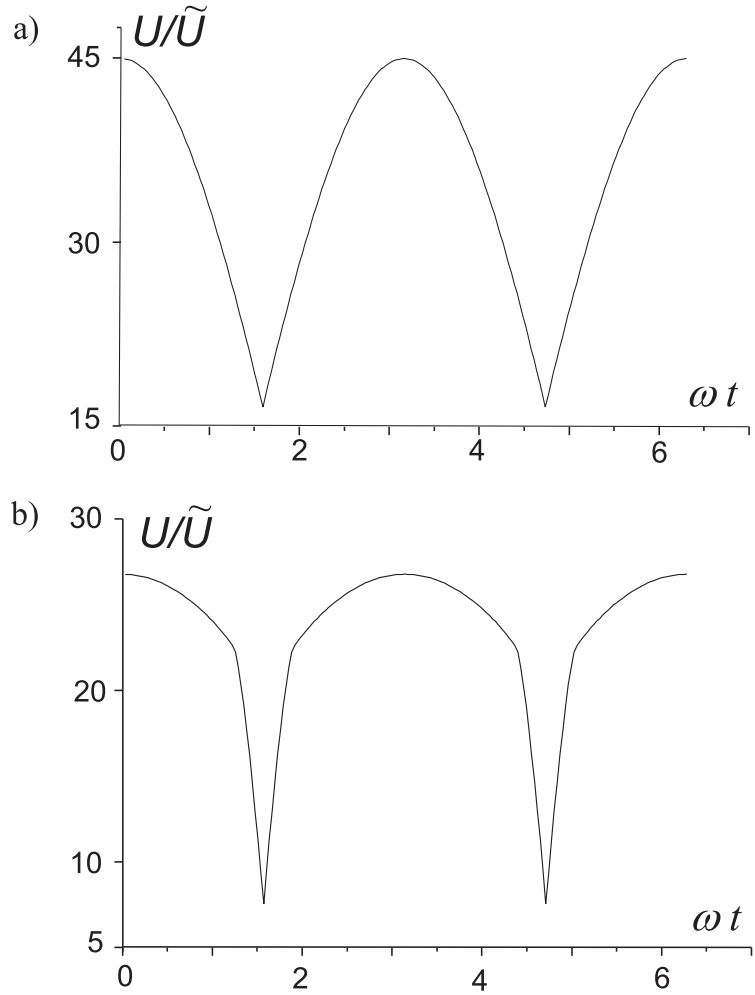}} \caption{Time dependence of the voltage $U$ at
relatively small (a, $h_m<H$) and large (b, $h_m>H$) AC
amplitudes.}
\end{figure}

The temporal dependence of voltage (\ref{VAX}) for the case when the
wave amplitude is not too large ($h_m<H$) and there exist the trapped
carriers during the whole period $2\pi/\omega$ is shown in Fig.~3,~a.
Fig.~3.~b represents the dependence $U(t)$ for the opposite case $h_m>H$,
in which during some part of the wave period (at $h_m|\cos\omega t|\geq H$)
the conductivity is caused by the flying particles only.

\section{Nonanalytical temporal dependence of electric field}

Knowing the vector potential $A(x,t)$, one can calculate
the rotational electric field ${\cal E}(x,t)$ as a correction to $E_0(t)$,
(see Ref.~(\ref{E-tot})). We are interested in the difference $\Delta
{\cal E}(t)={\cal E}(d/2,t)- {\cal E}(-d/2,t)$. This value is proportional
to the rate of alteration of the magnetic flux trough the cross-sectional
plane, which is perpendicular to the direction of the vector of the
total field ${\bf {\cal H}}(x,t)$,
and thus can be measured in experiment.

From Eqs.~(\ref{sol-1}) and (\ref{sol-2}), it follows
that the difference $a(1,t)-a(-1,t)$ is connected to the derivatives
$\partial a(1,t)/\partial \xi$
and $\partial a(-1,t)/\partial \xi$ by the relations,
\begin{eqnarray}
 a(1,t)-a(-1,t)=-\xi_0(t)\left[
\frac{\partial a(1,t)}{\partial \xi}-\frac{\partial a(-1,t)}{\partial \xi}
\right],\qquad h_m|\cos\omega t|\leq H,
\label{razn1}\\
 a(1,t)-a(-1,t)=
\frac{\partial a(1,t)}{\partial \xi}+\frac{\partial a(-1,t)}{\partial
\xi}, \qquad h_m|\cos\omega t|\geq H.
\label{razn2}
\end{eqnarray}
Let us now turn to the dimensional variables in Eqs.~(\ref{razn1}) and
(\ref{razn2}) using boundary conditions (\ref{cond-a}) and
relation (\ref{r}) between the values of $A({\rm sign}(\cos\omega t)
d/2,t)$ and $r$. After that one can obtain the following expression for the
magnitudes of the vector potential at the film boundaries:
$$
A({\rm sign}(\cos\omega t)d/2,t)=-{\tilde H}d\ln^2(R_+/d)r^2/4,
$$
\begin{equation}
A(-{\rm sign}(\cos\omega t)d/2,t)=-{\tilde H}d\ln^2(R_+/d)r^2/4
-2H|x_0(t)|
\label{Apm1}
\end{equation}
at
$$
h_m|\cos\omega t|\leq H
$$
and
\begin{equation}
A({\rm sign}(\cos\omega t)d/2,t)=0, \quad
A(-{\rm sign}(\cos\omega t)d/2,t)=-dh_m|\cos\omega t|
\label{Apm2}
\end{equation}
at
$$
h_m|\cos\omega t|\geq H.
$$
Formulae (\ref{Apm1}) and (\ref{Apm2}) are sewn at the time
moment when $h_m|\cos\omega t|=H$. The parameter $r$ in Eq.~(\ref{r})
vanishes, and the plane $x=x_0(t)$ coincides with one of the boundaries of
the sample, $|x_0(t)|=d/2$.  From relations (\ref{Apm1}) and (\ref{E-tot}),
by means of formula (\ref{xi-0}) for $\xi_0(t)$, we derive the expression
for the difference $\Delta {\cal E}(t)$ of magnitudes of the electric field
at the film boundaries,
\begin{equation}
 \Delta {\cal E}(t)=-\frac{2H}{c}\frac{\partial x_0(t)}{\partial t} =
-\frac{H(I)Lh_m}{2\pi}
\frac{\partial }{\partial t}
\left[\frac{\cos\omega t}{\sigma_{fl}(t)U(t)}\right],
\qquad h_m|\cos\omega t|\leq H.
\label{deltaE1}
\end{equation}
If the inequality $h_m\leq H$ holds, the previous relation
is valid during the whole period of the wave. However, in the case
$h_m>H$, there exists a time interval when the plane $x=x_0(t)$ of
alternation of the sign of the total magnetic field is absent. If such a
situation takes place one should use formula (\ref{Apm2}) in order to obtain
the dependence $\Delta {\cal E}(t)$. Finally we come to the result below,
\begin{equation}
\Delta {\cal E}(t)=\Delta {\cal E}_L\sin\omega t,\qquad
\Delta {\cal E}_L=d h_m \omega/c, \qquad h_m|\cos\omega t|\geq H.
\label{deltaE2}
\end{equation}
From this, it follows that the difference $\Delta {\cal E}(t)$ is a harmonic
function of time, i.e.  the response of the film on the external
electromagnetic excitation turns out to be linear if there are no
trapped electrons. It is obvious that formula (\ref{deltaE2}) also
describes the
dependence $\Delta {\cal E}(t)$ at small magnitudes of the current $I$
($H\ll{\tilde H}$), when the contribution of trapped particles to
the conductivity is negligible during the whole period of the wave. Then
the value $\Delta {\cal E}_L$ represents the amplitude of a linear
response.

\begin{figure}[bt]
\centering \scalebox{0.9}[0.9]{\includegraphics[bb=82 256 508
505]{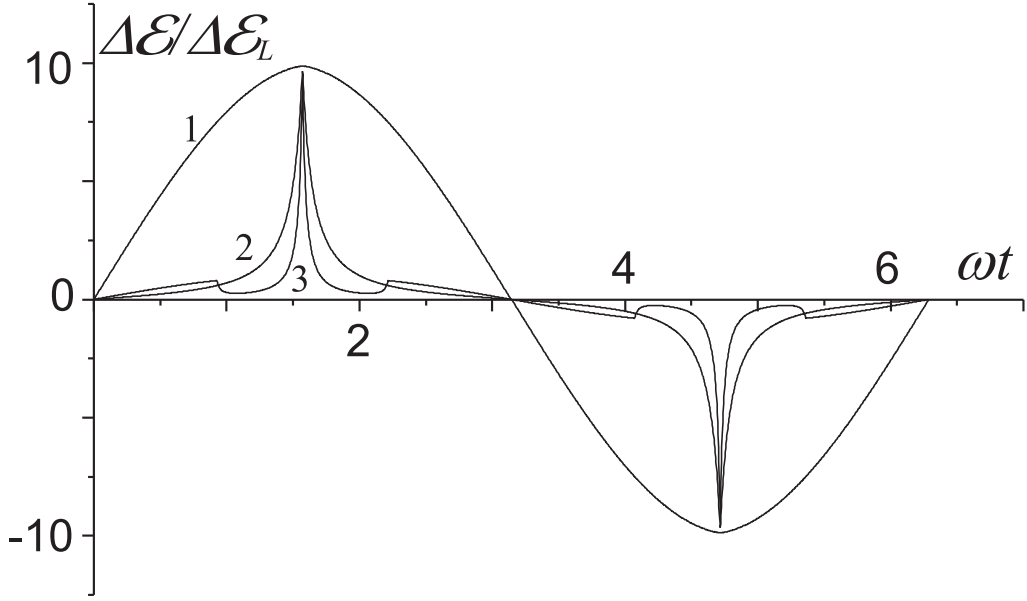}} \caption{The dependence $\Delta {\cal E}(t)$ for
$H=300{\tilde H}$ and various amplitudes of the AC signal:
$h_m=1{\tilde H}$ (1), $h_m=200{\tilde H}$ (2), $h_m=500{\tilde
H}$ (3).The ratio of the mean free path $l$ to the film thickness
$d$ equals to 30.}
\end{figure}
The dependence $\Delta {\cal E}(t)$ is shown in Fig.~4 for
a wide range of the AC amplitudes $h_m$ and for the
large magnitudes of the DC magnetic field $H$ of the current $I$,
when the inequality $H\gg {\tilde H}$ (or inequality (\ref{trap-cond}))
is valid. It is obvious that the ratio of the amplitude
$\Delta {\cal E}_m$ to its linear value $\Delta {\cal E}_L$
does not depend on $h_m$. From relations (\ref{deltaE1}), (\ref{VAX}),
and (\ref{eq-r}) at $\cos\omega t=0$, we find the expression
for $\Delta {\cal E}_m$,
\begin{equation}
 \frac{\Delta {\cal E}_m}{\Delta {\cal E}_L}=0.83
\left(\frac{H}{{\tilde H}}\right)^{1/2}\frac{1}{\ln(R/d)}, \qquad
\left(\frac{H}{{\tilde
H}}\right)^{1/2}\sim\frac{\sigma_{tr}}{\sigma_{fl}
}\Bigg|_{\cos\omega t=0}\sim\frac{l}{(Rd)^{1/2}}\gg 1.
\label{DeltaEm}
\end{equation}
The ratio $\Delta {\cal E}_m/\Delta {\cal E}_L$ is determined by the
magnitude of the DC magnetic field $H$ and can be
much greater than unity. In other words, there exists an effect of
amplification of the electric signal at the film surface. For small AC
amplitudes (curve $1$, $h_m=H/300$) the signal turns out to be
quasi-harmonic. However, with the increase of $h_m$ the dependence
$\Delta {\cal E}(t)$ shows kinks. Curve $2$ has kinks at
the points of extremum, i.e at the time moments when the AC magnetic
field $h_m\cos\omega t$ vanishes. These singularities
are related to the nonanalytical behavior of CVC of the film
(see. Eq.~(\ref{VAX}) and Fig.~3). Curve $3$ corresponds to the case
$h_m=5H/3$, in which the trapped electrons are absent during a part of
the wave period. In such a situation, the dependence $\Delta {\cal E}(t)$
contains additional kinks arising at the moments of appearance and
disappearance of the plane $x=x_0(t)$ of the sign alternation of the total
magnetic field.  They are located symmetrically with respect to the points
of extremum as shown in curve $3$. By means of formulae
(\ref{deltaE1}), (\ref{deltaE2}),(\ref{VAX}), and (\ref{eq-r}), we find
the right and left derivatives of the function $\Delta {\cal E}(t)$ at the
point $t_0=(1/\omega)\arccos(H/h_m)$ of the first kink,
\begin{equation}
\frac{\partial }{\partial t}\frac{\Delta{\cal E}(t)}
{\Delta {\cal E}_L}\Bigg|_{t=t_0-0}=\frac{\omega H}{h_m},
\label{der-left}
\end{equation}
\begin{equation}
\frac{\partial }{\partial t}\frac{\Delta{\cal E}(t)}
{\Delta {\cal E}_L}\Bigg|_{t=t_0+0}=\frac{\omega
H}{h_m}\left[1-\frac{\pi}{2\ln(R_+/d)}\left(\frac{H}{\tilde
H}\right)^{1/2}\left(\frac{h_m^2}{H^2}-1\right)\right].
\label{der-right}
\end{equation}
According to Eq.~(\ref{der-right}), the right derivative is negative and has
large absolute value even at $[(h_m/H)^2-1]\geq 1$.

\section{Surface impedance of film}

Let us analyze the dependence of the surface impedance at the film
boundary $x=d/2$ on the AC amplitude $h_m$ under conditions of interaction
of the transport current and the electromagnetic wave. The impedance
is proportional to the ratio of the first Fourier harmonics of the
electric ${\cal E}_\omega$ and magnetic $h_\omega$ fields at the surface of
the sample,
\begin{eqnarray}
 &Z=\frac{4\pi}{c}\frac{{\cal E}_\omega}{h_\omega}&=\frac{8\pi}{c}
\frac{{\cal E}_\omega}{h_m},\quad {\cal E}_\omega=
-\frac{\omega}{2\pi c}\int^{2\pi/\omega}_{0}\left(\frac{\partial
A(d/2,t)}{\partial t}- \frac{\partial \bar A}{\partial t}\right) e^{{\rm
i}\omega t} dt\nonumber\\
 &&=\frac{{\rm i}\omega^2}{2\pi c}\int^{2\pi/\omega}_{0}\left(A(d/2,t)-\bar
A(t)\right)e^{{\rm i}\omega t}dt.
\label{imped}
\end{eqnarray}
Taking into account Eqs.~(\ref{xi-0}) and (\ref{Apm1}), we deduce the
boundary value of the vector potential for the periods of
time given by the inequality $h_m|\cos\omega t|\leq H$,
\begin{equation}
 A(d/2,t)= \left\{
\begin{array}{ll}
-{\tilde H}d\ln^2(R_+/d)r^2/4,\,at \cos\omega t>0,&\\
-{\tilde H}\ln^2(R_+/d)r^2/4+cHLh_m\cos\omega t/2\pi U(t)\sigma_{fl}(t),
\,at \cos\omega t<0.&
\end{array}
\right.
\label{pot1}
\end{equation}

In the case $h_m|\cos\omega t|\geq H$, the following expression is valid
(see. Eq.~(\ref{Apm2})):
\begin{equation}
A(d/2,t)=\cases{0, &$\quad at \cos\omega t>0$,\cr dh_m\cos\omega t, &$
\quad at \cos\omega t<0$.}
\label{pot2}
\end{equation}
Let us calculate the mean value of the vector potential $\bar A(t)$
for $h_m|\cos\omega t|\leq H$, when there exists the plane of alternation
of sign of the field. According to Eqs.~(\ref{sol-1}) and (\ref{sol-2}),
we have
\begin{eqnarray}
\label{A1}
 &&\frac{\bar A(t)}{A({\rm sign}(\cos\omega t)d/2,
t)}=\frac{1}{2}\int^{1}_{-1}a(\xi,t)d\xi
=\xi_0(t)+(2u(t))^{1/2}(1+r(t)/3)^{1/2}
\xi^2_0(t)\nonumber\\
 &&+(2/3)u(t)\xi^3_0+\left(\frac{3}{4r(t)u(t)}\right)^{1/2}
\int^{1}_{0}\frac{\zeta d\zeta}{\sqrt{1-(1-\zeta)^{3/2}+3\zeta/2r(t)}}.
\end{eqnarray}
In the case $h_m|\cos\omega t|\geq H$, one should use solution
(\ref{sol-2}) with $r=0$, $\xi_* =1$ in order to find $\bar A(t)$.
Proceeding to dimensional variables and using Eqs.~(\ref{eps}) and (\ref{VAX1}),
one can easily obtain
\begin{equation}
\bar A(t)=-\frac{dh_m|\cos\omega t|}{2}-\frac{1}{6}Hd.
\label{A2}
\end{equation}
We draw reader's attention to the fact that the mean value of the
vector potential depends on time only via the term $|\cos\omega
t|$, $\bar A(t)=\bar A(|\cos\omega t|)$. This follows from
formulae (\ref{eq-r}), (\ref{eps}), and  (\ref{xi-0}) for the
values $r$ , $u$ , and $\xi_0$ as well as from the relation
(\ref{r}) between $A({\rm sign}(\cos\omega t)d/2, t)$ and $r$. It
also implies that the surface impedance in the main approximation
with respect to $d/\delta$ has imaginary part (reactance) only.
The latter is a consequence of the full transparency of the film.

We start calculation of the reactance $X$ with the case of
relatively small amplitudes $h_m<H$, when the group of trapped
electrons exists during the whole period of the wave. Let us
substitute expressions (\ref{pot1}) and (\ref{A1}) into
Eq.~(\ref{imped}). Then, the integrals containing $\bar A(t)$ and
$-{\tilde H}d\ln^2(R_+/d)r^2/4$ vanish since these functions
depend on $|\cos\omega t|$ only. By means of formula (\ref{VAX})
for the voltage $U$, the remaining integral can be transformed
into the form,
\begin{equation}
 X=\frac{8d\omega}{c^2}\int^{\pi/2}_{0}\frac{1+\bar
\sigma_{tr}(\tau)/\sigma_{fl}(\tau)}{1+(\bar\sigma_{tr}
(\tau)/\sigma_{fl}(\tau))(h_m/H)\cos\tau}\cos^2\tau d\tau,\quad h_m\leq H.
\label{small1}
\end{equation}
For the case of large amplitudes $h_m>H$, one should calculate the
reactance using formulae (\ref{pot1}), (\ref{pot2}), (\ref{A1}),
and (\ref{A2}). It represents a sum of two terms,
\begin{eqnarray}
 X&=&\frac{8d\omega}{c^2}\left[\int^{\pi/2}_{\pi/2-\arcsin
H/h_m}\frac{1+\bar \sigma_{tr}(\tau)/\sigma_{fl}}{1+(\bar\sigma_{tr}
(\tau)/\sigma_{fl}(\tau))(h_m/H)\cos\tau}\cos^2\tau d\tau\right.\nonumber\\
&+&\left.\int^{\pi/2-\arcsin H/h_m}_{0}\cos^2\tau d\tau\right],
\quad at\, h_m>H.
\label{big1}
\end{eqnarray}
The first term corresponds to the temporal interval when the trapped
electrons exist in the sample, and the second one is related to the
interval when these particles are absent.

Let us calculate the asymptotics of the surface reactance for the
case of rather large amplitudes $h_m\gg H$. For this purpose, we
rewrite integral (\ref{big1}) in another form,
\begin{eqnarray}
 X&=&\frac{8d\omega}{c^2}\left[\int^{\pi/2}_{0}\cos^2\tau
d\tau\right.\nonumber\\
 &+&\left.\int^{\pi/2}_{\pi/2-\arcsin
H/h_m}\left(\frac{1+\bar \sigma_{tr}(\tau)/\sigma_{fl}}{1+(\bar\sigma_{tr}
(\tau)/\sigma_{fl}(\tau))(h_m/H)\cos\tau}-1\right)\cos^2\tau
d\tau\right].
\label{big2}
\end{eqnarray}
In the second integral, we substitute the variable of integration
$(h_m\cos\tau)/H=\eta$ and expand the integrand in a power series in the
ratio $H/h_m$. Then one finds,
\begin{equation}
\frac{X}{X_L}=1+\frac{4}{\pi}(H/h_m)^3\int^{1}_{0}\left[\frac{1+\bar
\sigma_{tr} (\pi/2)/\sigma_{fl}(\pi/2)}{1+\bar \sigma_{tr}
(\pi/2)/\sigma_{fl}(\pi/2)\eta}-1\right]\eta^2d\eta, \label{big3}
\end{equation}
where
\begin{equation}
X_L=\frac{2\pi}{c^2}\omega d \label{linear}
\end{equation}
is the same as the value of reactance in the absence of the DC
transport current. The conductivities $\bar \sigma_{tr}(\pi/2)$,
and $\sigma_{fl}(\pi/2)$ are taken at the moment of time when the
AC magnetic field $h_m\cos\omega t$ turns into zero.  Therefore,
their ratio is much greater than unity due to inequality
(\ref{DeltaEm}). Taking into account condition (\ref{DeltaEm}), we
calculate integral (\ref{big3}) and obtain the following
asymptotics for the reactance,
\begin{equation}
\frac{X}{X_L}=1+\frac{2}{3\pi}\left(\frac{H}{h_m}\right)^3, \quad
H\ll h_m. \label{as-big}
\end{equation}

Now we consider the case of the extremely small amplitudes
described by the inequality $h_m\ll H\sigma_{fl}(\pi/2)/\sigma_{tr}(\pi/2)
\sim(H\tilde H)^{1/2}$. The integrand in Eq.~(\ref{small1})
can be presented as a power series in $h_m/(H\tilde H)^{1/2}$.
As a result the asymptotic takes the form,
\begin{eqnarray}
\frac{X}{X_L}&=&\frac{4}{\pi}\frac{\bar\sigma_{tr}(\pi/2)}{\sigma_{fl}(\pi/2)}
\int^{\pi/2}_{0}\left[1-\frac{\sigma_{tr}(\pi/2)}{\sigma_{fl}(\pi/2)}
\frac{h_m}{H}\cos\tau\right]\cos^2\tau d\tau\nonumber\\
&=&\frac{\sigma_{tr}(\pi/2)}{\sigma_{fl}(\pi/2)}(1-\frac{8}{3\pi}
\frac{\sigma_{tr}(\pi/2)}{\sigma_{fl}(\pi/2)}\frac{h_m}{H}), \quad
at \quad h_m\ll(H \tilde H)^{1/2}. \label{small-as}
\end{eqnarray}
We notice that reactance (\ref{small-as}) is
$\sigma_{tr}(\pi/2)\sigma_{fl}(\pi/2)\gg 1$ times greater than
that in the absence of the DC current. This is a direct
consequence of the effect of amplification of electric signal at
the film boundary which was treated in the previous section (see
Eq.~(\ref{DeltaEm})). The presence of strong DC current in the
sample also causes linear behaviour of the reactance in the region
of small amplitudes. As shown in Fig~5, the reactance decreases
monotonically within the region between asymptotics
(\ref{small-as}) and (\ref{as-big}).
\begin{figure}[bt]
\centering \scalebox{0.8}[0.8]{\includegraphics[bb=89 159 575
516]{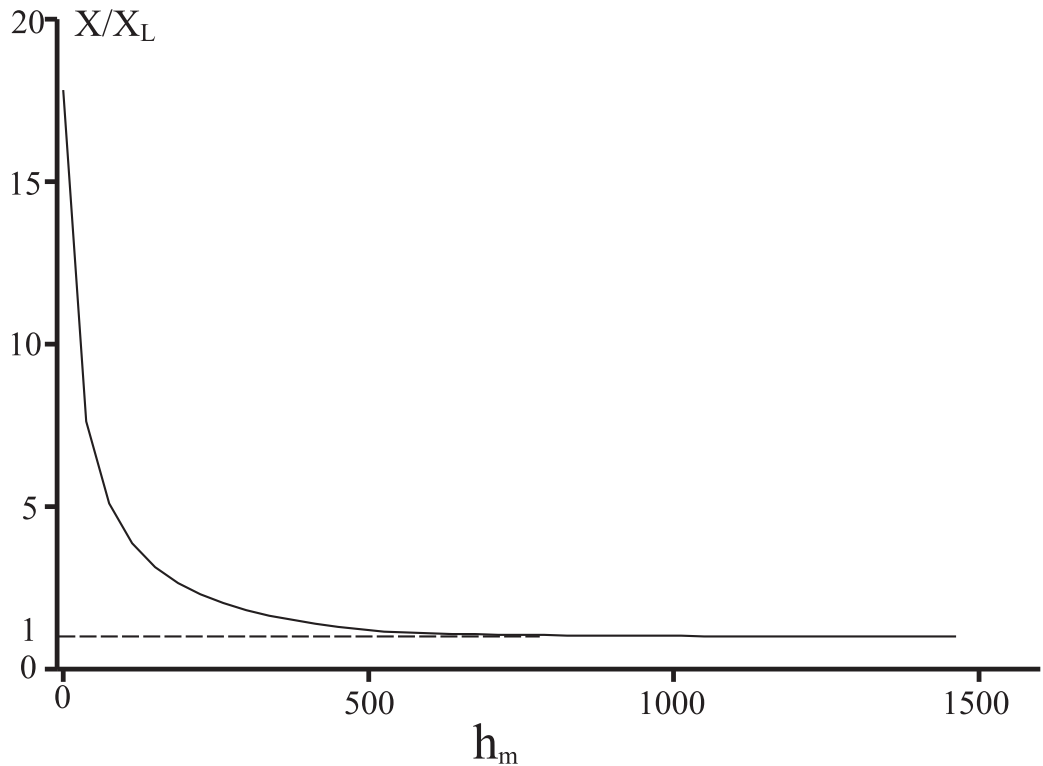}} \caption{The surface reactance $X$ (in units of the
reactance $X_L$ in the absence of the DC current) vs.
dimensionless amplitude $h_m/\tilde H$ of the AC signal at
$H=500{\tilde H}$.} \label{fig5}
\end{figure}

\section{Conclusion}
Nonlinear interaction of electromagnetic waves with a strong DC
transport current in a thin metal film leads to unusual physical
effects due to specific, typical only for metals, {\it
magnetodynamic} mechanism of nonlinearity. These effects have been
studied by analyzing the nonlinear response of the film, which
carries the DC current, irradiated bilaterally by electromagnetic
wave. The interaction of the wave with the current results in
nonanalytical behaviour of the AC electric field on the sample
surface which is characterized by appearance of sharp kinks. The
increase of the current is accompanied by a rise of the amplitude
of oscillations of the electric field at the surface of the
sample. This, in turn, causes to the growth of the imaginary part
of the surface impedance of the conductor.

The results obtained in this treatment are valid under certain applicability
conditions. Firstly, the AC electric field $\Delta {\cal E}(x,t)$ must be
small comparing to the potential electric field $E_0(t)$. It follows from
formulae (\ref{E-tot}), (\ref{Apm1}), and (\ref{Apm2}) that the quantities
${\cal E}$ and $\Delta {\cal E}_m$, Eq.(\ref{DeltaEm}), are of the same
order. Therefore, to ascertain the restrictions imposed by the condition
${\cal E}\ll E_0(t)$, we can use quantity $\Delta {\cal E}_m$ in the
latter condition. The quantity $\Delta {\cal E}_m$should be much less than
the minimum value the function $E_0(t)$, i.e. the magnitude of potential
field (\ref{VAX}) for $\cos\omega t=0$. The desired inequality reads
\begin{equation}
d^2\frac{h_ml}{HR}\ll\delta^2_n(\omega),\qquad
\delta^2_n(\omega)=\frac{c^2}{4\pi\sigma_0\omega},
\label{quasi1}
\end{equation}
where $\delta_n(\omega)$ represents the characteristic penetration depth
of the AC field into a metal under the condition of normal skin effect.
Secondly, the non-uniform component of
magnetic field inside the film must necessarily be much less than $h_m$.
This stems from the assumption that the AC magnetic field $h(x,t)$ should
be quasi-uniform ($h(x,t)\simeq h_m\cos\omega t$) across the bulk
of the film.
The maximum value of the non-uniform correction can be estimated from the
first of Maxwell's equations (\ref{Max}) as $(4\pi\sigma_{tr}\Delta {\cal
E}_md/c) \sim h_m(d/\delta)^2$, where an effective penetration depth
$\delta(\omega)$ equals to $\delta_n(\omega)(R/l)^{1/2}$. As a result, we
come to a requirement of the quasi-uniform property of the AC magnetic field
which can be written in the following form:
\begin{equation}
d^2\frac{l}{R}\ll\delta^2_n(\omega).
\label{quasi2}
\end{equation}
Comparing the restrictions imposed by inequalities (\ref{quasi1})
and (\ref{quasi2}), it can be
seen that condition (\ref{quasi1}) is more strict at large
AC amplitudes, $h_m>H$, while for small values of $h_m$ one
should use inequality (\ref{quasi2}).

For a sample with thickness
$d=10^{-3}$ cm, the electron free path $l=10^{-1}$ cm, the concentration of
electron $N=10^{23}$ cm$^{-3}$, the Fermi momentum
$p_F=10^{-19}$ g$\cdot$cm/sec and for magnetic fields $h_m=H=100$ Oe,
we have  $\omega<10^5$ sec$^{-1}$ using conditions (\ref{quasi1}) and
(\ref{quasi2}).
At such values of the parameters, conditions (\ref{quasi1}) and
(\ref{quasi2}) are fulfilled as well as condition
(\ref{trap-cond}), which states that the value of the mean free path
of electron should be large.

Unusual manifestation of specific magnetodynamic mechanism of
nonlinearity discussed in the present paper calls for future
investigation. In particular it would be very interesting to
explore experimentally theoretical predictions made in this work.

{}

\end{document}